\documentclass[aps,prb,twocolumn,showpacs,floatfix]{revtex4}

\usepackage{graphicx}% Include figure files
\usepackage{dcolumn}% Align table columns on decimal point
\usepackage[english]{babel}

\begin{document}

\title{Asymmetric $I$-$V$ characteristics and magnetoresistance in magnetic point contacts}
\author{A. R. Rocha and S. Sanvito}
\email{sanvitos@tcd.ie}
\affiliation{Department of Physics, Trinity College, Dublin 2, IRELAND}

\date{\today}

\begin{abstract}
We present a theoretical study of the transport properties of magnetic point contacts under bias.
Our calculations are based on the Keldish's non-equilibrium Green's function formalism combined with 
a self-consistent empirical tight-binding Hamiltonian, which describes both strong ferromagnetism and
charging effects. We demonstrate that large magnetoresistance solely due to electronic effects
can be found when a sharp domain wall forms inside a magnetic atomic-scale point contact. Moreover
we show that the symmetry of the $I$-$V$ characteristic depends on the position of the domain wall
in the constriction. In particular diode-like curves can arise when the domain wall is placed off-center
within the point contact, although the whole structure does not present any structural asymmetry.
\end{abstract}


\maketitle

\section{Introduction}\label{intro}

The study of electronic devices at the nanoscale is of paramount importance both from the scientific and the technological point of 
view. One important aspect is related to spin polarized electronic transport in magnetic materials, a field
usually known as spintronics \cite{Prinz}. This
has been subject of intense study in recent years, \cite{garcia1999,garcia2001,zchopra} since the discovery of the giant magnetoresistance
effect (GMR) in magnetic multilayers. \cite{baibich_gmr,binasch_gmr,pratt_gmr,gijs_gmr} GMR is the drastic change of the electrical 
resistance of a magnetic multilayer when a magnetic field is applied. The effect of the field is to
align ferromagnetically the magnetization vectors of adjacent magnetic layers, which otherwise are oriented 
antiparallel to each other. The parallel state has a lower resistance than the antiparallel
and this produces the magnetoresistance.
GMR has already led to the construction of the present generation of magnetic data storage
devices. However, in order to reach storage densities of the order of Terabit/in$^2$ a substantial downscaling 
of the read/write devices is needed. One possible avenue to this target is given by magnetic point contacts 
(MPC's), where the lateral size of the typical contacts approaches the atomic scale.

Recent experiments have shown that nanoscaled magnetic point contacts may present huge magnetoresistance
\cite{garcia1999,garcia2001,zchopra,versluijs,oscar1,oscar2,oscar3} reaching up a few hundred thousand percent. \cite{comment} 
This result alone can be seen as a major leap toward nanoscopic magnetic memory read/write devices. In addition, it is 
important to report that electro-deposited nickel point contacts present highly asymmetric $I$-$V$ characteristics typical of a diode-like 
behaviour. \cite{oscar1,oscar2} A complete explanation of both these effects is still lacking.

To date there is a large debate around the origin of the large magnetoresistance effects in MPC's. On the one hand, it has been argued
that magnetic field-induced mechanical effects can produce a large GMR. In fact, either magnetostriction, dipole-dipole 
interactions between the apexes\cite{chung_dipole} and magnetically induced stress relief\cite{ansermet} may have the effect of 
compressing the nanocontact once a magnetic field is applied. This enlarges the cross section of the MPC and consequently 
the resistance of the junction decreases. On the other hand, there are also strong indications that
mechanical effects alone are not able to account for whole magnetoresistance. Specifically Garc\'{\i}a {\it et. al.} 
have shown that the behaviour of MPC's does not comply with mechanical changes, in particular with 
magnetostriction effects. \cite{garcia1999,garcia2001}  

One important step toward the understanding of the magnetoresistance effect in point contacts comes from the work of 
Bruno.\cite{bruno} He has shown that if the magnetic moments on the apexes of the constriction are not aligned
parallel to each other, then a domain wall (DW) will be formed inside the constricted region. The main features of this 
DW are quite different from those of the well know Bloch \cite{bloch,landau} and N\'eel \cite{neel} walls. In particular,
the crucial point is that the length of the DW is predicted to be as long as the diameter of the constriction.
This result suggests that the DW trapped in an atomic scale MPC may be very sharp, only a few atomic 
planes long. At this length scales a DW produces a rather strong spin-dependent scattering potential and 
large magneto-induced effects are somehow expected. In this context it is important to report that
such an extreme confinement has been already achieved by contacting two magnetic grains with different orientation.
\cite{zchopra}

Since Bruno's seminal idea a number of theoretical works on transport through magnetic domain walls have been presented. 
\cite{maekawa,maekawa2000,nakamura,kelly} These consider equilibrium transport in the spirit of the Landauer-B\"uttiker 
formalism; \cite{landauer} current induced effects such as charging of the point contact and, quite possibly, 
movement of the domain wall have not been taken into consideration. However in low dimensional systems such 
as these even small biases might cause charge accumulation inside the MPC, changing its transport properties.\cite{ruitenbeek}

In this work we investigate whether the transport properties of MPCs, in particular the magnetoresistance effect and the 
asymmetry in the $I$-$V$ curves, can arise from electronic effects only. We model the magnetoresistance in MPC
using the typical spin-valve scheme: we assume that in the absence of a magnetic field, the magnetization vectors of 
the two leads are aligned opposite to each other (antiparallel state). Therefore in the zero field situation
a sharp DW is formed inside the MPC. Then, when a magnetic field is applied, the 
magnetic moments of the leads align parallel to each other and the wall is eliminated (parallel state). 
In this configuration the resistance is reduced giving rise to a GMR-like effect. Our task is to
calculate the $I$-$V$ curves of both the parallel and antiparallel states, assuming a simple but reasonable electronic
structure for the material forming the MPC and considering charging effects. 

We also investigate the effects of the position of the DW inside the constriction. When the DW
is positioned in the centre of the constriction, we expect symmetric $I$-$V$ curves, since no asymmetry is present
in the MPCs. However, if the position of the wall is shifted from the midpoint, and charging effects
are present, we expect the appearance of asymmetric $I$-$V$ curves. It is noteworthy to mention that a 
linear response Landauer-B\"uttiker approach to transport cannot account for charging effects, and therefore
cannot describe asymmetric $I$-$V$ characteristics.

This paper is organized as follows. In the next sections we introduce our self-consistent technique based on the non-equilibrium Green's 
function formalism \cite{keldysh,datta,Tsukada,stokbro} with tight-binding Hamiltonians.\cite{zhao} Then we will use our 
method to study the effects due to the existence of a domain wall on the electronic transport of MPC's.

\section{Non-equilibrium transport method}\label{theory}

In order to describe the transport properties of MPC's, including charging effects, we use the non-equilibrium Green's function approach
based on the Keldysh formalism.\cite{keldysh,datta,Tsukada,stokbro} We start by dividing an open system into 
three distinct regions: two semi-infinite electrodes and a central scattering region where the MPC is located. 
Here the leads act as charge reservoirs, therefore they set the temperature and electronic 
distribution of the junction in the steady state.
In principle calculating the electronic structure of such a system would involve the solution of a quantum mechanical problem 
for a non-periodic open system, which consist of an infinite number of degrees of freedom. However, there are two
reasons that allow us to focus only on the central scattering region. Firstly, we note that most of the potential drop
occurs at the nanoconstriction. This observation underpins the assumption that the relevant modification of the electronic 
structure due to the presence of a bias will occur to those degrees of freedom corresponding to the central region.
Secondly we assume that the rearrangement of the electronic structure in the central region does not
change the electronic structure of the electrodes. Note that this is a well sustained approximation for good metals where
the screening length is very short and the relaxation of the charge density occurs within a few atomic planes.
Under this condition their dynamics can be neglected with the exception of a rigid shift of the chemical potential due to 
the bias. Therefore we reduce our problem to that of calculating the Green's function of the central scattering region. 

Using an orthogonal tight-binding model we construct the retarded Green's operator of the scattering region in the
presence of the leads as follows\cite{datta}
\begin{equation}\label{green}
\mathbf{G}=\lim_{\eta\rightarrow 0} \left[\left(\epsilon+i\eta\right)-H_{\mathrm{S}}-\mathbf{\Sigma}_\mathrm{L}-
\mathbf{\Sigma}_\mathrm{R}\right]^{-1},
\end{equation}
where $H_\mathrm{S}$ is the Hamiltonian of the scattering region and
$\mathbf{\Sigma}_\mathrm{L}$ and $\mathbf{\Sigma}_\mathrm{R}$ are the leads self-energies. These are energy dependent potentials 
added to the Hamiltonian to account for the interaction with both the left and the right lead. The self-energies are 
matrices of the same dimension of $H_\mathrm{S}$ and of the form
\begin{equation}\label{selfenerg1}
\mathbf{\Sigma}_\mathrm{L}=H_\mathrm{LS}^\dagger\hat{g}_{\mathrm{L}}H_\mathrm{LS}
\end{equation}
and
\begin{equation}\label{selfenerg2}
\mathbf{\Sigma}_\mathrm{R}=H_\mathrm{RS}\hat{g}_{\mathrm{R}}H_\mathrm{RS}^\dagger,
\end{equation}
where $\hat{g}_\mathrm{L}$ and $\hat{g}_\mathrm{R}$ are the surface Green's functions for the left and right leads respectively and
$H_\mathrm{LS}$ ($H_\mathrm{RS}$) is the coupling matrix between the
scattering region and the left (right) lead. Note that $\mathbf{G}$ given by the equation (\ref{green}) is formally 
the Green's function of the effective Hamiltonian $H_\mathrm{eff}$ for the scattering region in presence of the leads
\begin{equation}\label{effH}
H_\mathrm{eff}=H_{\mathrm{S}}+\mathbf{\Sigma}_\mathrm{L}+\mathbf{\Sigma}_\mathrm{R}\:.
\end{equation}
This Hamiltonian is non-hermitian since the $\mathbf{\Sigma}$'s are not hermitian, consequently the number of electrons in the scattering
region is not conserved. 

One of the crucial aspects of the method is computing the surface Green's functions $\hat{g}$.
Several ways for calculating the surface Green's functions of a semi-infinite system 
are available in the literature, ranging from recursive methods \cite{datta} to semi-analytical solutions. \cite{sanvito}
In this work we use the technique presented by 
Sanvito {\it et. al.} \cite{sanvito} that calculates the surface Green's function exactly. 
The method first constructs the Green's function of an infinite system and then derives the surface Green's functions
$\hat{g}$ by introducing the appropriate boundary conditions. Following this procedure a general semi-analytical 
expression for the surface Green's function is derived. The benefit of our method is to avoid the typical 
complications of the recursive methods, preserving information about the individual scattering states in the 
leads. \cite{sanvito}

This formulation of the transport problem can account for finite bias.
As we discussed previously the bias $V$ is assumed not to change the electronic structure of the leads,
but only to shift all energy levels. Therefore $V$ is introduced as a rigid shift of the leads' Hamiltonian 
on-site energies
\begin{equation}
\mathbf{H}_\mathrm{L/R}\longrightarrow \mathbf{H}_\mathrm{L/R} \pm \frac{V}{2} {\cal I},
\end{equation}
where ${\cal I}$ is the identity matrix. This in turn redefines the leads self-energies in the following way
\begin{equation}
\mathbf{\Sigma}_\mathrm{L/R}\left(\varepsilon\right) \longrightarrow \mathbf{\Sigma}_\mathrm{L/R}\left(\varepsilon \mp V/2 \right).
\end{equation}

In contrast to the leads, due to the presence of a large potential drop, the effects of a finite bias on the scattering 
region need to be calculated self consistently. This is done by noting that $H_{\mathrm{S}}$ is a function of the 
scattering region single particle charge density only\cite{kohn} $H_\mathrm{S}=H_\mathrm{S}(\rho)$. This property gives 
us a clear prescription for performing a self-consistent calculation. 

We first compute the scattering region Green's function
(equation (\ref{green})) for $H_{\mathrm{S}}=H_{\mathrm{S}}(\rho_0)$, where the Hamiltonian $H_{\mathrm{S}}$ is 
evaluated at a given initial charge density $\rho_0$. Then from the
Green's function $\mathbf{G}$ we calculate the new charge density $\rho_1$, which is then used to construct the new Hamiltonian 
$H_{\mathrm{S}}\left(\rho_1\right)$. This procedure is iterated until reaching self-consistency, that is when 
$\rho_{\mathrm{n+1}}=\rho_\mathrm{n}$. Note that a fundamental requirement is that the final self-consistent density
matrix matches exactly that of the leads at the boundaries.
Therefore we usually enlarge the scattering region to contain a few atomic planes of the leads (those where the electronic 
structure is different from that of the bulk). The exact number of such planes depends on the screening length of the 
material considered. As a general rule we add as many planes as necessary for the charge density of the more external plane 
to match exactly that of the infinite leads. In this case the description of the leads in terms of self-energies 
is perfectly justified. 

Within this scheme the density operator of the scattering region $\mathbf{D}$ is evaluated from the Green's function $\mathbf{G}$
\begin{equation}\label{density}
\mathbf{D}=\frac{1}{2\pi}\int d\varepsilon \ \ \mathbf{G}\left[\mathbf{\Gamma}_\mathrm{L} f\left(\varepsilon-\mu_\mathrm{L}\right) +
\mathbf{\Gamma}_\mathrm{R}f\left(\varepsilon-\mu_\mathrm{R}\right)\right] \mathbf{G}^\dagger,
\end{equation}
where
\begin{equation}
\mathbf{\Gamma}_{\mathrm{L/R}}=i\left[\mathbf{\Sigma}_{\mathrm{L/R}}-\mathbf{\Sigma}_{\mathrm{L/R}}^\dagger\right],
\end{equation}
and $f\left(\epsilon-\mu_{\mathrm{L/R}}\right)$ is the Fermi distribution function calculated at the chemical potentials
of the leads $\mu_{\mathrm{L/R}}$.

In general it can be very clumsy to solve the equation (\ref{density}) numerically, since the integration has to be performed over
the entire real axis. However, we can both add and subtract the term $\mathbf{G}\mathbf{\Gamma}_\mathrm{R} \mathbf{G}^\dagger
f\left(\varepsilon-\mu_\mathrm{L}\right)$ to the equation (\ref{density}) rewriting the density operator as
\begin{equation}\label{DM}
\mathbf{D}=\mathbf{D}^0+\mathbf{D}^V,
\end{equation}
where now
\begin{equation}
\mathbf{D}^0=-\frac{1}{\pi}\int d\varepsilon \ \ \mathrm{Im}\left[ \mathbf{G}\right] f\left(\varepsilon-\mu_\mathrm{L}\right), 
\label{D_equil}
\end{equation}
\begin{equation}\label{D_out_equil}
\mathbf{D}^V= \frac{1}{2\pi}\int d\varepsilon \ \ \mathbf{G} \mathbf{\Gamma}_\mathrm{R} \mathbf{G}^\dagger
\left[ f\left(\varepsilon-\mu_\mathrm{R}\right) -
f\left(\varepsilon-\mu_\mathrm{L}\right)\right]. 
\end{equation}

We can interpret $\mathbf{D}^0$ as the density matrix at equilibrium, {\it i.e.} the one calculated when both the reservoirs have
the same chemical potential $\mu_\mathrm{L}$, and $\mathbf{D}^V$ as the out of equilibrium charge density obtained when a
potential bias $V$ is applied across the device. The separation of $\mathbf{D}$ into $\mathbf{D}^0$ and $\mathbf{D}^V$ has some important
numerical advantages. Firstly $\mathbf{D}^V$ is bounded by the difference between the
Fermi functions $f\left(\varepsilon-\mu_\mathrm{R}\right) -
f\left(\varepsilon-\mu_\mathrm{L}\right)$ and therefore one needs to perform the integration only in the energy range between the two 
chemical potentials. Secondly, since the Green's function $\mathbf{G}$ is analytical,
$\mathbf{D}^0$ can be integrated in the complex plane using a standard contour integral technique. \cite{lang,stokbro} 
In this way we have reduced an unbounded integral over the real energy axis to a bounded one over the real axis plus a well behaving 
unbounded integral in the complex plane. In addition, the lower limit of the contour integral is set by the lowest between the bottom 
of the conduction band of the leads and the deepest of the energy levels of the scattering region.

Once self-consistency has been achieved and both the charge density and the potential have been calculated, we can finally 
evaluate the current through the device
\begin{equation}
I=\frac{e}{h}\int d\varepsilon \ \ \mathrm{Tr}\left(\mathbf{\Gamma}_\mathrm{L} \mathbf{G}^\dagger \mathbf{\Gamma}_\mathrm{R} \mathbf{G} \right)
\left(f\left(\varepsilon-\mu_\mathrm{L}\right) - f\left(\varepsilon-\mu_\mathrm{R}\right)\right)\:.
\label{current}
\end{equation}

The term $\mathrm{Tr}\left(\mathbf{\Gamma}_\mathrm{L} \mathbf{G}^\dagger \mathbf{\Gamma}_\mathrm{R}\mathbf{G} \right)$ is the 
energy-dependent total transmission probability $T(\varepsilon, V).$\cite{datta} Note that the total current is nothing but
$e/h$ times the energy-dependent total transmission coefficient integrated over the bias window. This latter depends not only
on the electron energy but also over the bias applied $V$, since the potential across the device is bias-dependent.
However, this does not mean that the electrons exchange energy while passing through the device. On the contrary the transport
is still phase-coherent. The current is carried by electrons crossing the device with probability $T(\varepsilon,V)$,
leaving an occupied state of the lead with the higher chemical potential and filling an empty state at the same energy
in the other lead. At low bias ($\mu_\mathrm{R}\:\rightarrow\:\mu_\mathrm{L}$) and temperature ($T\:\rightarrow\:0$), 
the equation (\ref{current}) reduces to the well-known linear response Landauer-B\"uttiker formula.\cite{datta}

\section{The Model}

\begin{figure}[ht]
\includegraphics[width=6.5cm,clip=true]{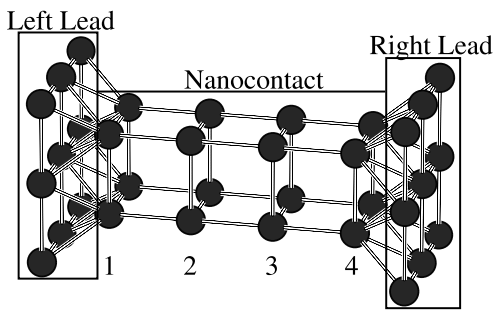}
\caption{Sketch of model structure used in the calculations consisting of two nine atom
cross section semi-infinite leads, connecting through a four atom cross section wire.
The numbers label the four planes of the wire.}
\label{drawing}
\end{figure}

The theoretical approach to transport described in the previous section is general and can be applied to 
a number of different systems and model Hamiltonians. Figure \ref{drawing} shows a sketch of the point contact 
used in the present calculations. Our structure consists of 
two semi-infinite rods representing the leads connected through a constricted region containing four atomic planes with 
a square four-atom cross section. This models a nickel nanowire connected to two current/voltage electrodes. 
Each atomic site is described by a spin-dependent two-orbital tight-binding model. One orbital forms a broad band and
represents the $s$ orbitals while the other forms a narrow band and represents the $d$ orbitals. 
Spin degeneracy is lifted on the $d$ states by adding a spin-dependent contribution to the on-site energy. The spin $\sigma$ on-site
energy for the $i$-th orbital at each site ($i$ labels both the position $\vec{R}$ and the orbital $\alpha$, $i=(\vec{R},\alpha)$) 
is given by
\begin{equation}\label{Escat}
\epsilon_i^\sigma=\epsilon_i^0+\frac{z_\sigma J}{2}\mu_{\vec{R}}\delta_{\vec{R}\alpha\:\vec{R}d}+
U\left(\sum_{\alpha\sigma}n^V_{\vec{R}\alpha\sigma}-n^0_{\vec{R}}\right).
\end{equation}
The first term, $\epsilon_i^0$ is the spin independent band centre. The second term, which is not zero only for the $d$ orbitals
is a Stoner-like term with $J$ the exchange integral, $z_{\uparrow\downarrow}=\mp 1$ and 
$\mu_{\vec{R}}=n_{\vec{R}d\uparrow}-n_{\vec{R}d\downarrow}$ 
the magnetic moment at each site.
This latter is given by the difference in the $d$ orbital occupation between majority and minority spins. Finally, the last term is 
introduced to ensure charge neutrality. \cite{vega} 
It is proportional to the difference between the self-consistent atomic charge density 
$n^V_{\vec{R}}=\sum_{\alpha\sigma}n^V_{\vec{R}\alpha\sigma}$ and that at equilibrium $n^0_{\vec{R}}$. 
$n^V_{\vec{R}}$ is calculated self-consistently from the non-equilibrium density matrix $\mathbf{D}$ 
(equation (\ref{DM})). The total number of electrons and the orbital occupation are respectively given by
\begin{equation}
N_{e}=\mathrm{Tr}\left[\mathbf{D}\right]
\end{equation}
and
\begin{equation}
n_{\vec{R}\alpha\sigma}^V=\mathrm{Tr}[\mathbf{D}|\vec{R}\alpha\sigma\rangle\langle \vec{R}\alpha\sigma|]\:,
\end{equation}
where $|\vec{R}\alpha\sigma\rangle\langle \vec{R}\alpha\sigma|$ is the projector over the state 
$(\vec{R},\alpha,\sigma)$.
Finally here we consider only first nearest neighbour coupling with elements $t_s$, $t_d$ and an 
hybridization term $t_{sd}$. 

\begin{figure}[ht]
\includegraphics[width=7.5cm,clip=true]{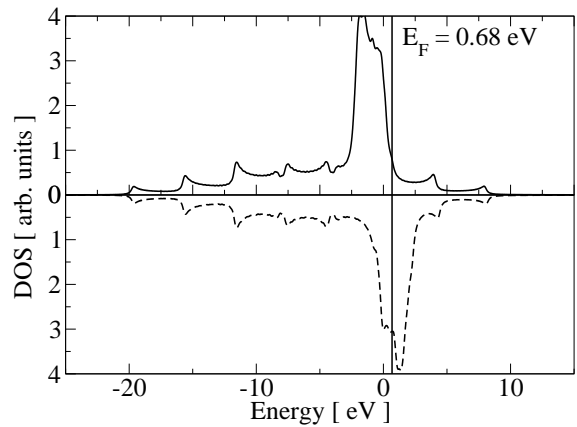}
\caption{Density of states of a simple cubic infinite rod described by our self consistent two bands tight-binding model. 
The picture refers to the same nine atom cross section structure used as semi-infinite lead in the transport calculation.
The solid (dashed) line corresponds to majority (minority) spin band.}
\label{DOS}
\end{figure}

Before turning our attention to the non-equilibrium problem, we must set the parameters of the calculation, 
{\it i. e.} we must determine some reasonable values for $\epsilon_i^0$, $t_i$, $J$ and $U$. 
In particular, the leads depicted in figure \ref{drawing} should describe the bulk properties of a strong ferromagnet 
(a ferromagnet where the majority spin band is completely occupied). 
To this purpose we performed self-consistent calculations of the electronic structure of a 3$\times$3 infinite rod 
(representing the leads). We assume an atomic occupation of three electrons per atom (two electrons in the 
$s$ orbitals and one electron in the  $d$ orbital), and we fix $t_s$ and $t_d$ in order to reproduce the correct
bandwidth of the $s$ and $d$ band of Ni respectively. \cite{me2} Then we set $J$, $t_{sd}$ and $U$ to give 
respectively the Stoner instability, the charge transfer from the $d$ to the $s$ band, and the local 
charge neutrality. A good description of a strong ferromagnet is given by the following
choice of parameters: $\epsilon_s=-6$~eV, $\epsilon_d=0$~eV, $t_s=-2.8$~eV, $t_d=-0.3$~eV, 
$t_{sd}=0.4$~eV, $U=7$~eV and $J=3.6$~eV. Figure \ref{DOS} shows the DOS resulting from our 
calculations, where one can clearly see a completely filled $d$ band for majority spin and an 
half-filled $d$ band for the minority. 

We are now ready to calculate the $I$-$V$ curve of the nanocontact.
In our calculations, we considered two possible configurations depending on whether the magnetizations in each of the leads are
aligned parallel or anti-parallel with respect to each other. In the latter case, a domain wall will appear in the middle of the
constriction with characteristic length scales comparable to the length of the nanowire. In addition, 
we study the effect of the position of the DW on the $I$-$V$ characteristics of the MPC. In all cases presented here we consider 
collinear spins. This is justified by a recent calculation from Imamura {\it et. al.} \cite{maekawa} who have used a Heisenberg 
model in a mean field approximation to show that, in the case of a domain wall pinned in a constriction, there is no spin 
precession and minority and majority bands can be treated separately.

\section{Results and Discussion}

As mentioned previously we model a point contact following the sketch of figure \ref{drawing}, where we consider a simple cubic
lattice of nickel atoms with the MPC composed by a two-atom-wide four-atom-long chain connected to two semi-infinite rods
representing the current/voltage electrodes.

We start by considering the situation where no magnetic field is applied to the MPC. In this case we assume that 
a DW forms inside the constriction. However it is worth noting that the position of the DW is not {\it a priori} determined, since 
in general it is given by pinning centers in the junction. These are not present in our calculations 
and therefore we have considered two possibilities: 1) the DW is situated half way across the device 
(symmetrical case), and 2) the DW is situated between the third and fourth transversal plane
(asymmetrical case). It important to stress that in both the situations the domain wall is abrupt
and that the direction of the magnetization changes by 180$^o$ across the Bloch wall.

\begin{figure}[h]
\center
\includegraphics[width=7.0cm,clip=true]{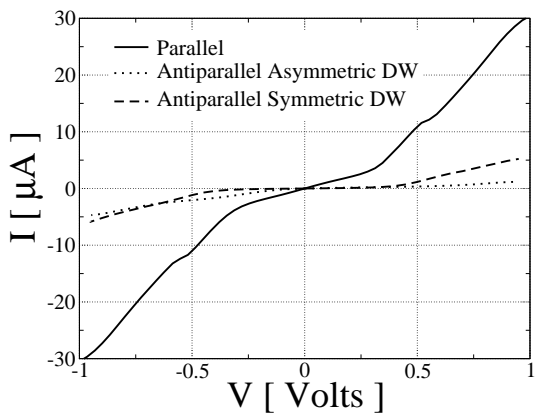}
\caption{Current as a function of bias applied for both parallel (solid line) and antiparallel
alignment of the magnetizations in the leads. For the antiparallel configuration we consider
the domain wall positioned either between the second and the third plane of the MPC (symmetric DW, dashed line)
and between the third and the fourth plane (asymmetric DW, dotted line). Note the large magnetoresistance
in both the cases and the large asymmetry of the $I$-$V$ characteristic for the asymmetric domain
wall.}\label{asymmetric}
\end{figure}

Figure \ref{asymmetric} shows our results for all these cases. As a matter of notation, in all the $I$-$V$ curves 
presented hereafter we define 
positive bias when the current flows from the left hand-side to the right hand-side (from plane one to four) and we use the left-hand
side lead to set the spin direction. In figure \ref{asymmetric} the dashed curve shows our result for a symmetric DW and the
dotted line the asymmetric. We can clearly see that an asymmetric $I$-$V$ curve appears when the domain wall is asymmetrically 
placed with respect to the junction. 

\begin{figure}[h]
\center
\includegraphics[width=7.507cm,clip=true]{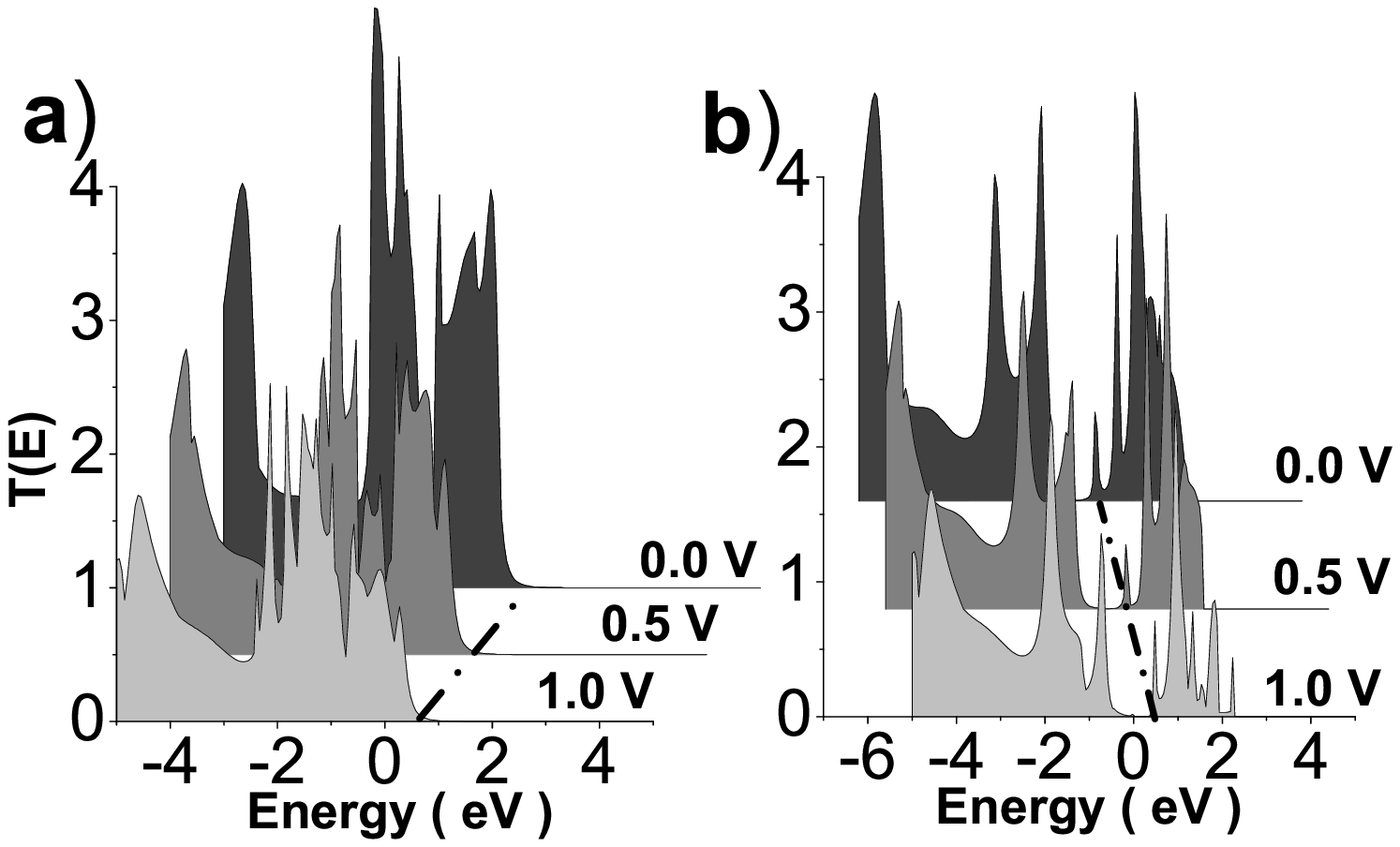}
\includegraphics[width=7.500cm,clip=true]{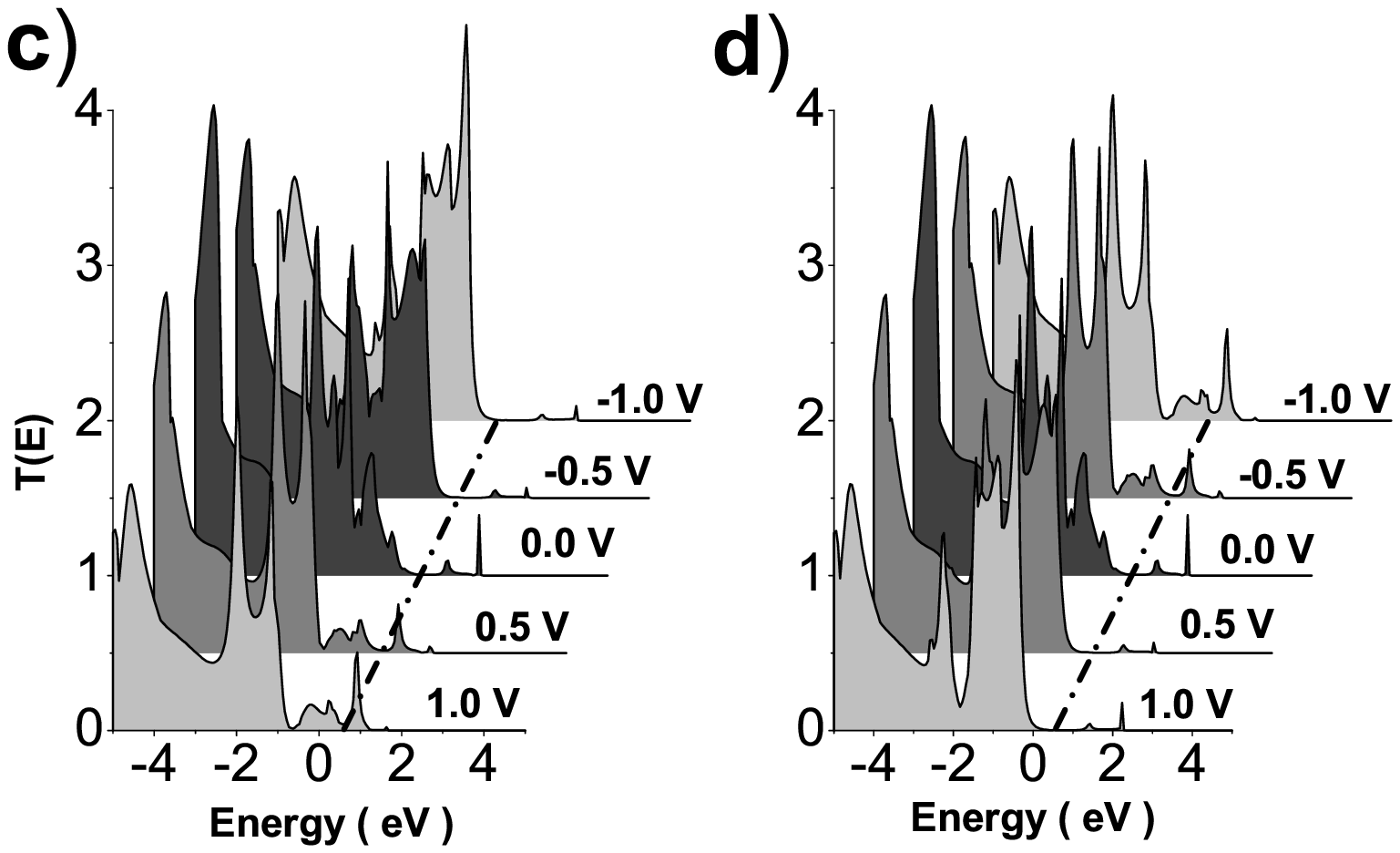}
\includegraphics[width=7.5cm,clip=true]{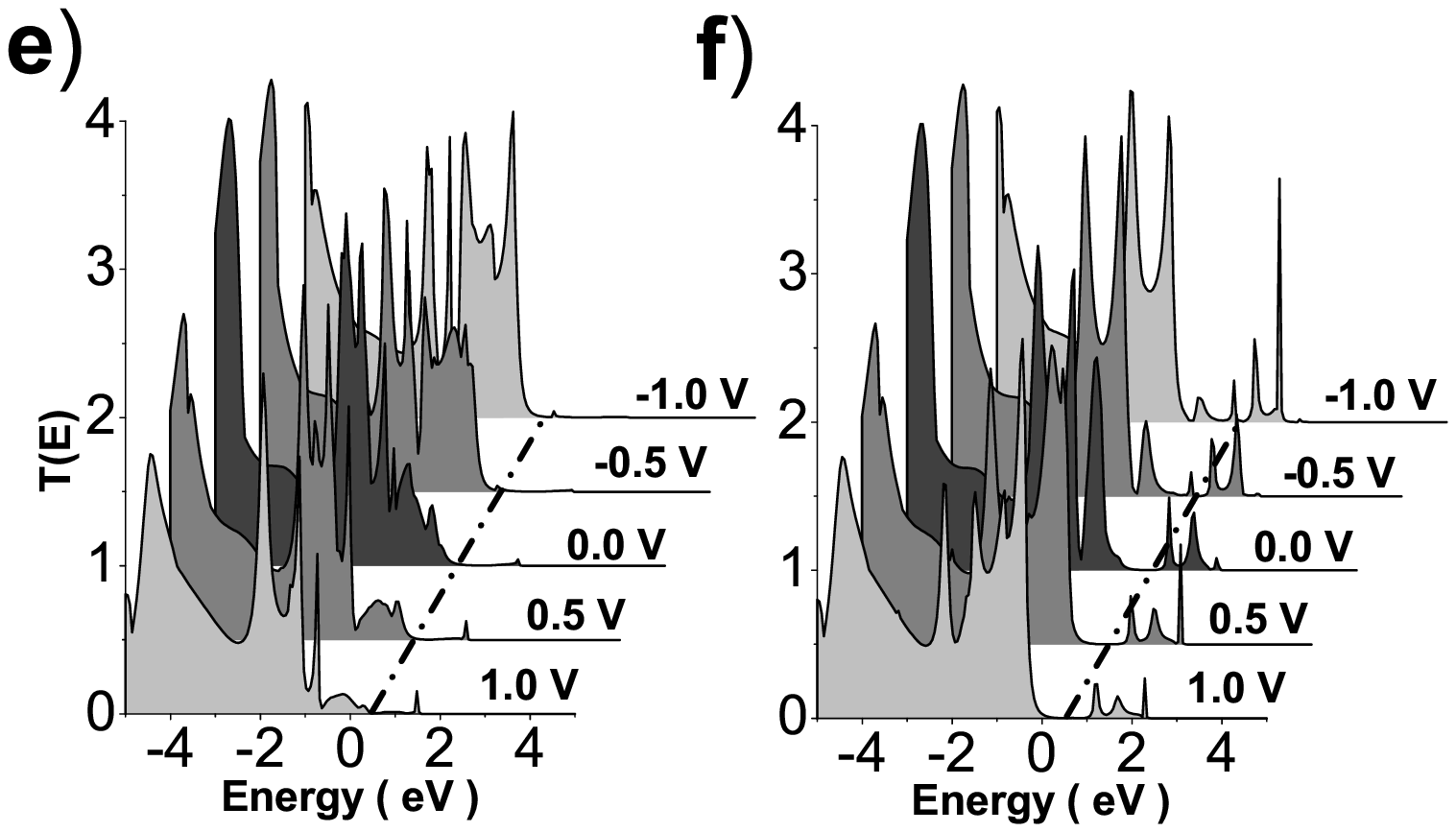}
\caption{Transmission coefficient as a function of energy for different biases and spin. The
different curves refer to: 
a) parallel case majority spin,
b) parallel case minority spin,
c) antiparallel symmetric case majority spin, 
d) antiparallel symmetric case minority spin,
e) antiparallel asymmetric case majority spin,
f) antiparallel asymmetric case minority spin. 
We define positive bias when the current is flowing from the 
left to the right, and the spin direction with respect to the
magnetization direction of the left hand-side lead. The dashed line denotes the position of the
Fermi level ($E_\mathrm{F}=0.68$~eV).}\label{transm}
\end{figure}

Before discussing the reason for this asymmetry we consider the case when a magnetic field strong enough to align
the magnetic moment of the leads is applied. We simulate this situation by setting the magnetization of the leads 
to be parallel to each other. The self consistent solution now does not present any Bloch wall and the corresponding $I$-$V$
characteristic is also shown in figure \ref{asymmetric} (solid line). From the picture it is clear that the 
current for the parallel configuration is larger than that of both the antiparallel ones at any bias. This of course
means that the point contact presents positive magnetoresistance at any bias. It is also worth noting that
such a magnetoresistance increases as the bias increases, in particular for the case in which 
the DW is asymmetrically placed in the junction. This suggests that a large GMR 
entirely due to the electronic properties of the point contact can be obtained, although for a fully
quantitative estimation a more realistic description of the electronic structure and the structural
conformation of the point contact is needed.

Let us now turn our attention to the asymmetry properties of the $I$-$V$ curves. In order to achieve a
qualitative understanding of the underling physics it is useful to calculate the transmission
coefficients as a function of the energy for different biases and for the three situations
studied (parallel, antiparallel symmetry and antiparallel asymmetric). These are presented in figure \ref{transm}. 

Since the current is essentially given by the energy integral of the transmission coefficient
over the bias window (see the equation (\ref{current})), from the picture one can establish
the relative contribution of the different spin currents to the total current for a given bias.
For the parallel case (figure \ref{transm}a and \ref{transm}b) the current has contributions 
from both the spin directions, although the majority spins contribute to the transmission coefficient 
below the Fermi level and the minority contribute above. Borrowing the notation from molecular
transport theory \cite{datta} we can say that the majority spin conductance is through the
highest occupied molecular orbital (HOMO), while the minority is through the lowest unoccupied
molecular orbital (LUMO). 

\begin{figure}[h]
\center
\includegraphics[width=7.5cm,clip=true]{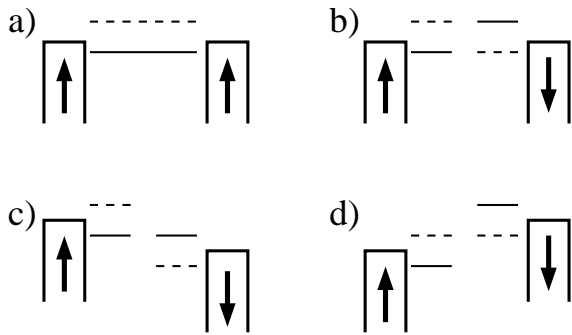}
\caption{Cartoon showing the levels alignment in the magnetic point contact. The solid 
(dashed) line denotes a majority (minority) spin molecular state. a) parallel case, b)
antiparallel symmetric case at zero bias, c) antiparallel symmetric case at positive bias,
and d) antiparallel symmetric case at negative bias. Note that for the antiparallel case 
the spin of the resonant level is opposite for opposite bias direction.}\label{cartoon}
\end{figure}

The situation changes drastically when we look at the antiparallel alignment. Consider first
the symmetric DW case. From the figures \ref{transm}c and \ref{transm}d one can see that the
conductance around $E_F$ is dominated by majority spins for positive bias and by minority spins
for negative bias. We can understand this behaviour by looking at the cartoon of figure 
\ref{cartoon}.

Here we schematically model our magnetic point contact as a magnetic molecule. When the magnetization
of the two leads are aligned parallel to each other (and to the molecule) the molecule presents
a majority spin HOMO and a minority spin LUMO state (figure \ref{cartoon}a). Both these states
extend through the whole molecule and can give rise to resonant transport. 

In contrast, in the 
antiparallel case the DW formed inside the molecule splits the HOMO and LUMO states. In fact,
since the transmission through the DW is small and the coupling with the leads strong, we can now model
our system as two molecules strongly attached respectively to the left and the right lead, but weakly 
coupled to each other.
This gives rise to the level scheme presented in figure \ref{cartoon}b, which is strictly valid
only in the case the DW resistance is infinite. Within this scheme the left hand-side part of
the PC couples strongly with the left lead therefore presenting a majority spin HOMO and a minority spin LUMO.
The situation is opposite on the right hand-side since the magnetization of the right lead is
rotated. Recalling the fact that we do not consider the possibility of spin mixing, this configuration presents
a large resistance since there are no resonant states with the same spin extending through
the entire point contact.
If we now apply a bias there will be level shifting. This aligns the majority spins HOMO on the left
with the LUMO on the right for positive bias (figure \ref{cartoon}c) and the minority LUMO on the left
with the HOMO on the right for negative bias (figure \ref{cartoon}d). Therefore this mechanism
leads to majority spin conductance for positive bias and minority spin conductance for negative.

\begin{figure}[ht]
\center
\includegraphics[width=7.5cm,clip=true]{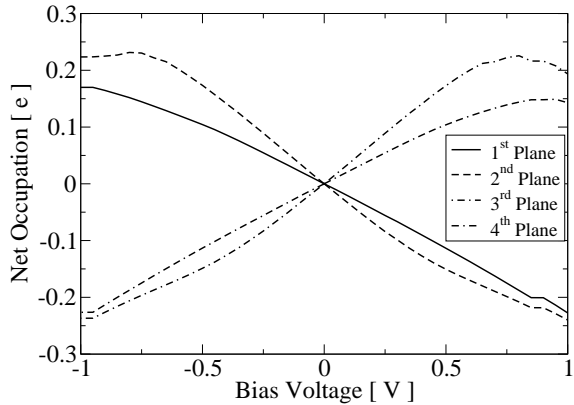}
\caption{Net charge accumulated in the point contact as a function of bias and the atomic position: antiparallel 
symmetric case. The net charge is calculated as the occupation difference per plane between the self-consistent charge
and that obtained at zero bias. Positive (negative) sign indicates that electrons have been removed from 
(donated to) the plane.}\label{occup_sym}
\end{figure}

From this simple scheme it is also easy to explain the symmetry properties of the $I$-$V$ curves. 
These are mainly given by the charging properties of the whole structure. For positive bias
and symmetric DW charge is accumulated in the first two atomics planes to the left of the DW and 
it is depleted in the two atomic planes to the right. This is demonstrated in figure \ref{occup_sym}
where we present the net charge on each atomic plane as a function of bias voltage. According to 
the equation (\ref{Escat}) charge accumulation (depletion) moves the molecular levels toward 
higher (lower) energies. In the case of a symmetric DW the molecular states to the left and 
to the right of the Bloch wall charge in a symmetric way with respect to the bias direction. This
means that the level alignment responsible for the large current will occur for the same bias
difference independently from the bias polarity.

\begin{figure}[h]
\center
\includegraphics[width=7.0cm,clip=true]{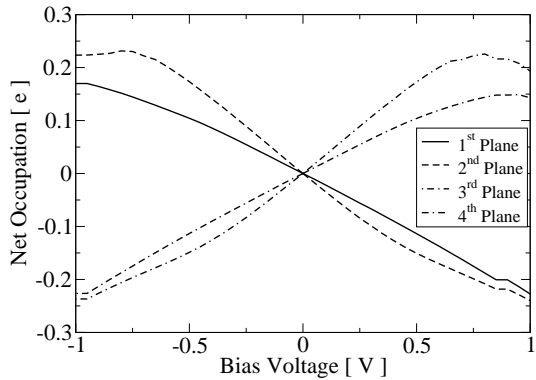}
\caption{Net charge accumulated in the point contact as a function of bias and the atomic position: antiparallel 
asymmetric case. The domain wall is positioned between the third and fourth plane. The net charge is 
calculated as the occupation difference per plane between the self-consistent charge
and that obtained at zero bias. Positive (negative) sign indicates that electrons have been removed from 
(donated to) the plane.}
\label{occup_asym}
\end{figure}

In contrast, when the DW is between the third and the fourth plane, more charge can be
accumulated (depleted) to the left of the DW with respect to the right (see figure \ref{occup_asym}),
since more states (atomic planes) are available. 
In other words the energy levels do not shift at the same rate for positive and negative bias.
This means that the alignment of the energy level now depends
on the bias polarity, leading to an asymmetric $I$-$V$ curve. Note that the mechanism for asymmetry
proposed here does not require different coupling between the point contact and the two leads,
but only a non-symmetric position of the DW within the point contact.

In summary, we have calculated the $I$-$V$ curve for a model magnetic point contact, and demonstrated that 
large magnetoresistance can be obtained solely from its electronic properties. This is the result of 
strong scattering through a domain wall pinned inside the contact. We have also investigated the r\^ole
of the position of the DW inside the junction and its effect on the $I$-$V$ curves. Our main result is that
largely asymmetric $I$-$V$ curves can be found when the DW is asymmetrically placed inside the point contact,
although the whole structure does not present any structural asymmetry. We have interpreted this result
in terms of the charging properties of the junction and of spin-dependent HOMO/LUMO alignment.
Although our results are semi-quantitative and a more accurate description of the electronic structure 
of the materials forming the point contact is needed, we believe that large asymmetry of the $I$-$V$ 
curve and GMR are a common feature of atomic scale magnetic point contacts.

\section*{Acknowledgments}

This work is sponsored by Science Foundation of Ireland under the grant SFI02/IN1/I175.
A.R.R. thanks Enterprise Ireland for financial support (grant EI-SC/2002/10).
We acknowledge useful discussions with J.M.D.~Coey and O.~C\'espedes~Boldoba.
%\bibliography{nickelbib,greenfbib,qconducmetalbib}

\begin{thebibliography}{10}
\bibitem{comment}
\bibinfo{note}{Throughout the paper we will use the ``optimistic'' definition of magnetoresistance
$\mathrm{MR}=\frac{R(0)-R(B)}{R(B)}$, where $R(B)$ is the magnetic field dependent resistance of the 
device}.
\end{thebibliography}

\end{document}